\documentclass[preprint,preprintnumbers,amsmath,amssymb]{revtex4}

\usepackage{epsfig}
\usepackage{hyperref}

\begin{document}
\title{The Cross Section of the Process
 $e^{+}+e^{-}\rightarrow J/\psi+\eta_c$ within the QCD Light-Cone Sum Rules}
\author{Yan-Jun Sun$^{1,\;2}$\footnote{Email: sunyj@ihep.ac.cn}, Xing-Gang
Wu$^3$\footnote{Email: wuxg@cqu.edu.cn}, Fen Zuo$^4$\footnote{Email:
zuofen@itp.ac.cn}and  Tao Huang$^1$ \footnote{Email:
huangtao@ihep.ac.cn}}
\address{$^1$Department of Modern Physics, University of Science and
Technology of China, Hefei 230026, P.R. China\\
$^2$Institute of High Energy Physics and Theoretical Physics Enter
for Science Facilities,Chinese Academy of Sciences,
Beijing 100049, P.R. China\\
$^3$Department of Physics, Chongqing University, Chongqing 400044,
P.R. China\\
$^4$Key Laboratory of Frontiers in Theoretical Physics, Institute of
Theoretical Physics, Chinese Academy of Sciences, Beijing 100190,
P.R. China}

\date{\today}

\begin{abstract}
We calculate the cross section of the exclusive process $e^{+}
+e^{-}\rightarrow J/\psi+\eta_c$ at the leading order approximation
within the QCD light-cone sum rules approach. It is found that the
form factor $F_{VP}(V=J/\psi,P=\eta_c)$ depends mainly on the
behavior of the twist-2 distribution amplitude of the $\eta_c$-meson
at the scale of this process. Thus in order to obtain a reliable
estimation of the cross section, it is important to have a realistic
distribution amplitude of the $\eta_c$ meson, and to deal with the
evolution of the distribution amplitude to the effective energy
scale of the process. Our results show that one can obtain a
compatible prediction with the Belle and BaBar experimental data. \\


\end{abstract}

\maketitle

\section{introduction}
The double-charmonium production in $e^{+}e^{-}$ annihilation at the
B factory provides a good platform to study both perturbative and
non-perturbative effects in quantum chromodynamics (QCD).

On one hand, from the experimental point of view, the cross section
of the precess $e^{+}+e^{-}\rightarrow J/\psi+\eta_c$ was measured
at Belle \cite{Belle2002,Belle2004} and BaBar \cite{BaBar2005}, and
their recent observation show that \cite{Belle2002,BaBar2005}
\begin{displaymath}
\sigma(e^{+}+e^{-}\rightarrow J/\psi+\eta_c)\times{{\cal B}_{>2}} =
25.6\pm2.8\pm3.4 \; {\rm fb} \;\;({\rm Belle})
\end{displaymath}
and
\begin{displaymath}
\sigma(e^{+}+e^{-}\rightarrow J/\psi+\eta_c)\times{{\cal B}_{>2}}=
17.6\pm2.8^{+1.5}_{-2.1} \; {\rm fb} \;\;({\rm BaBar}),
\end{displaymath}
where ${{\cal B}_{>2}}$ denotes the branching fraction for the final
states with more than two charged tracks.

On the other hand, from the theoretical point of view, the process
$e^{+}+e^{-}\rightarrow J/\psi+\eta_c$ is usually studied within the
nonrelativistic QCD (NRQCD) \cite{BBL1997}. Under the leading order
(LO) NRQCD calculation, Refs.\cite{BLB2003,BL2003,Chao2003} derived
a much smaller cross section $\sigma=3.78 \sim 5.5 \;{\rm fb}$ for
the first time. Thus there are large discrepancy between theoretical
predictions of the LO NRQCD calculation and the experimental
measurements. In order to solve this problem, many attempts were
made based on the NRQCD approach. Refs.\cite{Chao2006,Wang2008}
diminished the disagreement to a large degree by including the
radiative correction. Moreover, as pointed out in
Refs.\cite{BKKY2007,Chao2007,BLC2008}, the relativistic corrections
can further improve the accuracy. By taking both the radiative and
relativistic corrections into account, Ref.\cite{BLC2008} got
$17.6^{+8.1}_{-6.7} \; {\rm fb}$, then the authors there
optimistically concluded that the disagreement between theoretical
predictions in NRQCD approach and experiments has been resolved
 \cite{BLC2008}. However, one may doubt the validity of the
$\alpha_s$- and $v^2$- expansion in the process
$e^{+}+e^{-}\rightarrow J/\psi+\eta_c$, if the LO result is an order
of magnitude smaller than the experimental measurements and the
next-to-leading order (NLO) corrections/higher $v^2$-expansion terms
inversely play a dominate role for the double charmonium production.
Furthermore, Ref.\cite{Wang2008} showed that the scale dependence of
the cross section can not be improved even with the NLO correction,
so it is an important matter to determine the typical scale of the
process or at least to make a more reliable estimation of scale
dependence.

In contrast, it was argued that the experimental results by Belle
and BaBar collaborations of the process $e^{+}+e^{-}\rightarrow
J/\psi+\eta_c$ can also be explained by using the perturbative QCD
(pQCD) with proper models for the charmonium distribution amplitudes
(DA) \cite{BC2005}. They claimed, ``the difficulties in explaining
the Belle and BaBar results for $\sigma(e^{+}+e^{-}\rightarrow
J/\psi+\eta_c)$ are not really the difficulties of QCD, but are
rather due to a poor approximation of the real dynamics of
$c$-quarks by NRQCD". Actually, the exclusive process can be
factorized into two parts in the pQCD approach: the calculable
hard-parton amplitude and the hadronic distribution amplitude. If
one replaces all DAs in the pQCD formulae by a simple $\delta$
function, the calculated cross section will be back to a few $fb$
that is consistent with the NRQCD approach. However one always
assume that the hadronic distribution amplitude is not
non-relativistic. Therefore, the key point is that the relativistic
DA, instead of the $\delta$ function, enhances the cross section of
the process $e^{+}+e^{-}\rightarrow J/\psi+\eta_c$.

Furthermore, since the QCD light-cone sum rules (LCSR) combines the
QCD sum rules and the pQCD theory of hard exclusive processes in a
suitable way, it can be a good tool for calculating the form factors
in the large momentum transfer. For example, LCSR is a successful
method for dealing with the $\gamma^{*}\rho\rightarrow \pi$
transition form factor \cite{K1999} that is similar to the process
$e^{+}+e^{-}\rightarrow J/\psi+\eta_c$. We shall try to apply the
QCD LCSR approach to calculate the amplitude of the process
$e^{+}+e^{-}\rightarrow V+P$, where $V=J/\psi,...$ and
$P=\eta_c,...$. The theoretical predictions at the LO approximation
for the cross section of the process $e^{+}+e^{-}\rightarrow V+P$ at
the large energy scale can be obtained by using the QCD LCSR.
Similar to the pQCD approach, this method also faces two problems:
which charmonium DA model should be adopted and how large effects
can be determined by the DA evolution. In the present paper, we
shall discuss the behavior of the different models for the
charmonium DA and the effects of the renormalized group evolution
with the effective scale in the process.

The remaining parts of the paper are organized as follows. In Sec.
II, we present calculation technology for $e^{+}+e^{-}\rightarrow
J/\psi+\eta_c$ under the QCD LCSR, where the leading-twist DA is
constructed and its QCD evolution is presented. Numerical results
for the cross section of the process $e^{+}+e^{-}\rightarrow
J/\psi+\eta_c$ are presented in Sec. III. The final section is
reserved for summary and conclusion.

\section{Calculation technology for $e^{+}+e^{-}\rightarrow J/\psi+\eta_c$}

Generally, the cross section for the process
$a(p_1)+b(p_2)\rightarrow c(p_3)+d(p_4)$ is given by
\begin{eqnarray}
\sigma=\frac{1}{4E_1E_2 v_{rel}}\int\frac{d^3 \vec{p}_3 d^3
\vec{p}_4} {(2\pi)^3 2E_3(2\pi)^3 2E_4}(2\pi)^4 \delta^4
(p_1+p_2-p_3-p_4)|\overline{\cal M}|^2,
\end{eqnarray}
where $p_i=(E_i,\vec{p}_i)$ $(i=1,\cdots,4)$ stand for the
four-momentum of initial and final particles correspondingly,
$v_{rel}=|\frac{\vec{p_1}}{E_1}-\frac{\vec{p_2}}{E_2}|$.
$|\overline{\cal M}|^2$ is the squared absolute value of the matrix
element, where the color states and spin projections of the initial
and final particles have been summed up and those of the initial
particles have been averaged.

For the exclusive double-charmonium production
$e^{+}+e^{-}\rightarrow J/\psi+\eta_c$, its Lorentz-invariant matrix
element turns out to be
\begin{equation}
{\cal M}=i\int d^4 x{<}J/\psi\eta_c|T\left\{Q_c J^{c}_{\mu}(x)
A^{\mu}(x),\bar{e}(0)\gamma_{\nu}e(0)A^{\nu}(0)\right\} |e^{+}e^{-}>
,
\end{equation}
where $J^{c}_{\mu}(x)=\bar{C}(x)\gamma_\mu C(x)$ is the $c$-quark
electromagnetic current. Then, we obtain
\begin{equation}
|\overline{\cal M}|^2 = 2Q_c^2  |F_{VP}|^2
\frac{(S-4m_{J/\psi}^2)}{4S}\left[1+\cos^2\theta\right],
\end{equation}
where $\theta$ is the scattering angle, $Q_c=\frac{2}{3}$ is the
charm quark charge and the form factor $F_{VP}$ is defined as
\begin{eqnarray}
{<}J/\psi(P-q)\eta_c(P)|J^c_\mu(0)|0{>}= \epsilon_{\mu
abc}\epsilon^{a *} q^b P^c F_{VP},\label{eq:fvpdef}
\end{eqnarray}
with $\epsilon^{a}$ being the polarization vector of $J/\psi$-meson
and $S=-q^2$. Neglecting the small mass difference between $J/\psi$
and $\eta_c$ mesons, the cross section becomes
\begin{eqnarray}
\sigma&=&\frac{\pi \alpha^2 Q_c^2
}{6}\left(1-\frac{4m_{J/\psi}^2}{S}\right)^{3/2} |F_{VP}|^2.
\end{eqnarray}

It is shown that the main part is to calculate the form factor
$F_{VP}$. There are many methods to calculate it, such as NRQCD
 \cite{BLB2003,BL2003,Chao2003,Wang2008,Chao2006,BKKY2007,Chao2007,BLC2008},
PQCD \cite{BC2005} and light-cone perturbative QCD approaches
\cite{CJ2007,Braguta2008}. Here we use the LCSR approach
\cite{BBK1989,BF1989,CZ1990,CK2001} to calculate $F_{VP}$.

\subsection{the form factor $F_{VP}$ within the QCD LCSR}

We adopt the following two-point correlator to calculate the form
factor $F_{VP}$
\begin{eqnarray}
\Pi_{\mu\nu}(P,q)&=& i \int d^4 xe^{-iqx}{<}\eta_c(P)|T\{J^c_\mu(x)
J^c_\nu(0)\}|0{>}, \label{eq:pimunu}
\end{eqnarray}
where $q$ is the four-momentum of the virtual photon, $P$ is the
four-momentum of $\eta_c$ meson.

On one hand, by inserting a complete set of intermediate hadronic
states in Eq.(\ref{eq:pimunu}), we get
\begin{eqnarray}
\Pi_{\mu\nu}(P,q)= -\epsilon_{\mu\nu\alpha\beta}q^\alpha P^\beta
F_{VP} f_{J/\psi}
\frac{1}{m_{J/\psi}^2-(q-P)^2}+\frac{1}{\pi}\int_{s_0}^{\infty} ds
\frac{{\rm Im} F_{\mu\nu}}{s-(q-P)^2} \label{eq:completation} ,
\end{eqnarray}
where the decay constant $f_{J/\psi}$ is defined as, $
{<}0|J^c_\mu(0)|J/\psi(P-q){>} = f_{J/\psi}
m_{J/\psi}\epsilon_{\mu}$, with $\epsilon^{\mu}$ being the
polarization vector of $J/\psi$-meson. $s_0$ is the threshold
parameter whose value can be taken as $3.6^2~\mbox
{GeV}^2<s_0<4.2^2~\mbox {GeV}^2$ \cite{EJ2001}. The second term in
Eq.(\ref{eq:completation}) is the dispersion integral that includes
the contributions from the excited and continuum states in the
region $s>s_0$.

On the other hand, the correlation function Eq.(\ref{eq:pimunu}) can
also be calculated by expanding the $T$-product of quark currents
near the light cone $x^2=0$ due to sufficiently large momentum
transfer. For such purpose, we contract the two $c$-quark fields and
write down a free $c$-quark propagator
\begin{eqnarray}
\overline{C(x)\bar{C}(0)}&=& i S(x,0)=i \int \frac{d^4
k}{(2\pi)^4}e^{-ikx}\frac{\not\!k+m_c}{k^2-m_c^2}.
\end{eqnarray}
Then up to twist-3 accuracy, Eq.(\ref{eq:pimunu}) can be simplified
as
\begin{eqnarray}
\Pi_{\mu\nu}^{QCD}(P,q)&=&2\epsilon_{\mu\rho\nu\tau}q^{\rho}
P^{\tau} f_{\eta_c} \int_{0}^{1}dx \frac{\phi_{\eta_c}(x)}{m_c^2-(x
P-q)^2},\label{eq:piQCD}
\end{eqnarray}
where $m_c$ is the current $c$-quark mass, $f_{\eta_c}$ is the decay
constant of ${\eta_c}$ meson, $\phi_{\eta_c}$ stands for the
$\eta_c$ leading-twist DA that is defined through the matrix
element:
\begin{eqnarray}
{<}\eta_c(P)|\bar{C}(x) \gamma^{\tau}\gamma_5
C(0)|0{>}&=&-iP^{\tau}f_{\eta_c} \int_0^1 du
e^{iuPx}\phi_{\eta_c}(u)+{\rm higher\; twist\; terms}.
\end{eqnarray}

Next, by applying the quark-hadron duality to
Eq.(\ref{eq:completation}) and by applying the Borel transformation
\cite{SVZ1979}
\begin{eqnarray}
{\cal B}_{M^2}\frac{1}{m^2_{J/\psi}-(q-P)^2}&=&\frac{1}{M^2}
e^{-\frac{m^2_{J/\psi}}{M^2}} , \nonumber\\
{\cal B}_{M^2}\frac{1}{m^2_c-(q-xP)^2},
&=&\frac{1}{xM^2}e^{\left\{-\frac{1}{xM^2}
\left[m^2_c+x(1-x)P^2-(1-x)q^2\right]\right\}},
\end{eqnarray}
to Eq.(\ref{eq:completation}) and Eq.(\ref{eq:piQCD}), we obtain the
sum rule for $F_{VP}$
\begin{eqnarray}
F_{VP}&=&\frac{2f_{\eta_c}}{m_{J/\psi} f_{J/\psi}} \int_\Delta^1 dx
\frac{\phi_{\eta_c}(x)}{x}e^{\left \{-\frac{1}{xM^2}\left[m_c^2
+x(1-x)m_{\eta_c}^2+(1-x)Q^2\right]
+\frac{m_{J/\psi}^2}{M^2}\right\}}, \label{eq:fvp}
\end{eqnarray}
where $\Delta =\frac{1}{2m_{\eta_c}^2}
\left[\sqrt{(s_0-m_{\eta_c}^2+Q^2)^2 + 4(m_c^2+Q^2)
m_{\eta_c}^2}-(s_0-m_{\eta_c}^2-q^2)\right] \label{eq:delta}$, $M^2$
is the Borel transformation parameter and $P^2 =m_{\eta_c}^2$,
$-q^2=Q^2=S=112\mbox {GeV}^2$. It is found that the form factor
$F_{VP}$ depends heavily on the DA $\phi_{\eta_c}$, especially on
its end point behavior due to $\Delta\sim 0.9$.

\subsection{leading-twist DA of $\eta_c$ meson}

The key input for the form factor is the gauge-independent and
process-independent DA $\phi_{\eta_c}$, which is of non-perturbative
nature and can be defined as the integral of the valence Fock wave
function \cite{LB1980}
\begin{eqnarray}
\phi_{\eta_c}(x,\mu_0)&=&\frac{2\sqrt{6}}{f_{\eta_c}}
\int_{|\vec{k_\perp}|^2<\mu_0^2}\frac{d^2
\vec{k_\perp}}{16\pi^3}\Psi_{\eta_c}(x,\vec{k}_\perp),\label{eq:mu}
\end{eqnarray}
where $\mu_0$ stands for the separation scale between the
perturbative and non-perturbative regions. As for a scale
$\mu>\mu_0$, the non-perturbative DA $\phi_{\eta_c}(x,\mu)$ is given
by the renormalization group evolution that can be calculated
perturbatively.

\begin{table}
\caption{The explicit form of the spin-space wave function
$\chi^{\lambda_{1}\lambda_{2}}(x,\vec{k}_\perp)$, where the
transverse momentum $\vec{k}_\perp=(k_x,k_y)$ and $m_c^{*}$ stands
for the constituent $c$-quark mass. }
\begin{center}
\begin{tabular}{|c||c|c|c|c|}
\hline\hline ~~~$\lambda_1\lambda_2$~~~ & ~~~$\uparrow\uparrow$~~~&
~~~$\uparrow\downarrow$~~~&
~~~$\downarrow\uparrow$~~~ & ~~~$\downarrow\downarrow$~~~ \\
\hline $\chi^{\lambda_{1}\lambda_{2}}(x,\vec{k}_\perp)$ &
$-\frac{k_x- i k_y} {\sqrt{2(m^{*2}_c+k_\perp^2)}}$ &
$\frac{m^*_c}{\sqrt{2(m^{*2}_c+k_\perp^2)}}$&
$-\frac{m^*_c}{\sqrt{2(m^{*2}_c+k_\perp^2)}}$ & $-\frac{k_x+i
k_y}{\sqrt{2(m^{*2}_c+k_\perp^2)}}$
\\\hline\hline
\end{tabular}
\label{tab0}
\end{center}
\end{table}

Up to now, it is difficult to give the light-cone wave function
(LCWF) from the first principles of QCD. So one usually constructs
some phenomenological models for the wave function, such as BC model
 \cite{BC2005}, BKL model \cite{BKL2006}, BLL model \cite{BLL2007}, MS
model \cite{MS2004}, BHL model \cite{BHL1981} and etc. Here, we
shall take the BHL model for the $\eta_c$ wave function, which can
be written as \cite{HZ2007}
\begin{equation}\label{eq:BHL}
\Psi^{\lambda_{1}\lambda_{2}}_{\eta_c}(x,\vec{k}_\perp)=\varphi_{\mathrm{BHL}}(x,\vec{k}_\perp)
\chi^{\lambda_{1}\lambda_{2}}(x,\vec{k}_\perp)=A
e^{-b^2\frac{\vec{k_\perp}^2+m_c^{*2}}
{x(1-x)}}\chi^{\lambda_{1}\lambda_{2}}(x,\vec{k}_\perp),
\end{equation}
where $m_c^{*}$ stands for the constituent $c$-quark mass,
$\lambda_1$ and $\lambda_2$ are helicity states of the constitute
$c$ and $\bar{c}$ quarks,
$\chi^{\lambda_{1}\lambda_{2}}(x,\vec{k}_\perp)$ stands for the
spin-space wave function coming from the Wigner-Melosh rotation
\cite{M1974}. $\chi^{\lambda_{1}\lambda_{2}}(x,\vec{k}_\perp)$ can
be found in Refs.\cite{spin1,spin2,spin3}, whose explicit form is
shown in TAB.\ref{tab0}.

\begin{figure}
\includegraphics[scale=1.20]{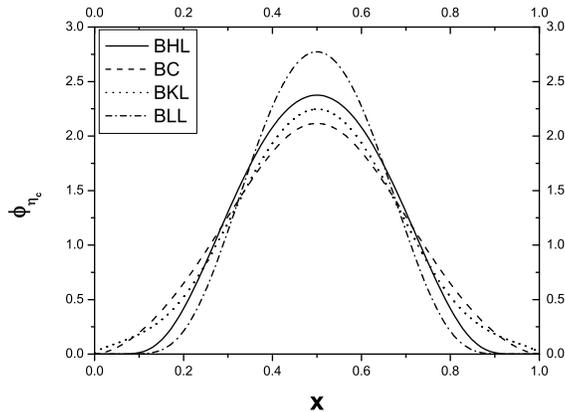}
\caption{The comparison of the $\eta_c$ DAs of BHL model with those
of BC \protect \cite{BC2005}, BKL \protect \cite{BKL2006}, BLL
\protect \cite{BLL2007} and BHL models at the initial scale $\mu_0$.
} \label{fig1}
\end{figure}

The parameters $A$ and $b^2$ can be determined by two constraints.
One constraint is from the wave function normalization
\begin{equation}
\frac{2\sqrt{6}}{f_{\eta_c}}\int_0^1 dx
\int_{|\vec{k_\perp}|^2<\mu_0^2}\frac{d^2
\vec{k_\perp}}{16\pi^3}\sum_{\lambda_1\lambda_2}
\Psi^{\lambda_1\lambda_2}_{\eta_c}(x_i,\vec{k}_\perp)=1.
\label{eq:leptonic}
\end{equation}
It can be found that only the usual helicity components
$\lambda_1+\lambda_2=0$ makes contribution to the wave function
normalization. More explicitly, from TAB.\ref{tab0}, we have
$\chi^{\lambda_1+\lambda_2=0}(x,\vec{k}_\perp)=\frac{Am_c^{*}}
{\sqrt{k^2_\perp+m_c^{*2}}}$. Another constraint is from the
probability of finding the leading Fock state $|c\bar{c}>$ in the
$\eta_c$ Fock state expansion, i.e.
\begin{equation}
\int_0^1 dx \int\frac{d^2 \vec{k_\perp}}{16\pi^3}
|\varphi_{\mathrm{BHL}}(x,\vec{k}_\perp)|^2 =P_{\eta_c},
\end{equation}
with $P_{\eta_c}\simeq 0.8$ \cite{spin1}. One can assume
$\mu_0=m_c^{*}$ to be the initial scale for the non-perturbative
distribution amplitude of the $\eta_c$-meson. Inputting the
constituent quark mass $m_c^{*}=1.8~\mbox {GeV}$ \cite{BC2005}, the
decay constant $f_{\eta_c}=0.335~\mbox {GeV}$ \cite{Yao2006} and the
initial scale $\mu_0=m_c^{*}=1.8~\mbox {GeV}$, we get the
corresponding parameters for $A=285.64291~\mbox
{GeV}^{-1},b^2=0.19057~\mbox {GeV}^{-2}$. We compare the $\eta_c$-DA
of our BHL model at the scale $\mu_0$ with those of BC
\cite{BC2005}, BKL \cite{BKL2006}, BLL \cite{BLL2007} models in
Fig.(\ref{fig1}).

However, the scale $\mu$ of the form factor (\ref{eq:fvp}) that is
at the B-factory is very different from the initial scale $\mu_0$ of
$\eta_c$ DA. When DA runs to a higher energy scale $\mu$, other than
$\mu_0$, with a proper QCD evolution, the behavior of DA shall be
changed to a certain degree, especially in its upper end-point
regions $\Delta<x<1$ that determines the form factor $F_{VP}$ as
shown by Eq.(\ref{eq:fvp}). Therefore, it is quite important to do
the DA evolution from the initial scale $\mu_0$ to a typical energy
scale $\mu$ of the process so as to derive a more reliable cross
section for $e^{+}+e^{-}\to J/\psi+\eta_c$ process. Thus, the next
section is devoted to deal with the evolution of the $\eta_c$ DA.

\subsection{The evolution of the $\eta_c$ DA with the scale $\mu$}

We describe the DA evolution according to Ref.\cite{LB1980}. In the
light-cone gauge, the DA $\phi_{\eta_c}$ is related to the hadronic
wave function $\Psi_{\eta_c}$, which is the Fourier transform of the
positive-energy projection of the usual Bethe-Salpeter wave function
evaluated at relative ``light-cone time", i.e.
\begin{eqnarray}
\phi_{\eta_c}(x_i,\mu)&=&\left(\ln
\frac{\mu^2}{\Lambda^2}\right)^{-\gamma_F/\beta}
\int_{|\vec{k_\perp}|^2<\mu^2}\frac{d^2
\vec{k_\perp}}{16\pi^3}\Psi_{\eta_c}(x_i,\vec{k}_\perp),\label{eq:BL}
\end{eqnarray}
 where the factor in front
of the integral comes from the scale dependence due to vertex and
self-energy insertions. An evolution equation is obtained by
differentiating both sides of Eq.(\ref{eq:BL}) with respect to
$\mu^2$. To order ${\cal O}(\alpha_s)$, we obtain an ``evolution
equation"
 \cite{LB1980}
\begin{eqnarray}
x_1x_2\mu^2\frac{\partial\,
\tilde{\phi}_{\eta_c}(x_i,\mu)}{\partial\,
\mu^2}=C_F\frac{\alpha_s(\mu^2)}{4\pi}
\left\{\int_0^1[d\,y]V(x_i,y_i)\tilde{\phi}_{\eta_c}(y_i,\mu)-
x_1x_2\tilde{\phi}_{\eta_c}(x_i,\mu)\right\},\label{eq:evolution}
\end{eqnarray}
where
\begin{eqnarray}
V(x_i,y_i)& =
&2C_F\left[x_1y_2\theta(y_1-x_1)\left(\delta_{h_1\bar{h_2}}
+\frac{\Delta}{(y_1-x_1)}\right)+(1\leftrightarrow2)\right], \\
\, [d\,y]&=&d\,y_1 d\,y_2\delta(1-y_1-y_2),\nonumber \\
\phi_{\eta_c}(x_i,\mu)&=&x_1 x_2
\tilde{\phi}_{\eta_c}(x_i,\mu),\nonumber
\end{eqnarray}
$C_F=4/3$, $\delta_{h_1\bar{h_2}}=1$ when the $c$ and $\bar{c}$
helicities are opposite, and $\Delta\tilde{\phi}_{\eta_c}(y_i,\mu)=
\tilde{\phi}_{\eta_c}(y_i,\mu)-\tilde{\phi}_{\eta_c}(x_i,\mu)$. The
running coupling constant $\alpha_s(\mu^2)$ at the LO is given by
$\alpha_s(\mu^2)=\frac{4\pi}{b_0\ln\left(\frac{\mu^2}
{\Lambda^2}\right)}$ with $b_0=25/3$. One explicit solution of
 Eq.(\ref{eq:evolution}) can be written in the following Gegenbauer
expansion
\begin{eqnarray}
\phi_{\eta_c}(x_i,\mu)&=&x_1x_2 \sum_{n=0}^{\infty}
a_n\left(\ln\frac{\mu^2}{\Lambda^2}\right)^{-\gamma_n}C^{3/2}_n(x_1-x_2),\label{eq:geg}
\end{eqnarray}
where the Gegenbauer polynomials $C^{3/2}_n$ are eigenfunctions of
$V(x_i,y_i)$ and the corresponding eigenvalues are the
``non-singlet" anomalous dimensions
\begin{eqnarray}
\gamma_n&=&\frac{C_F}{\beta}\left(1+4\sum_{k=2}^{n+1}\frac{1}{k}
-\frac{2\delta_{h_1\bar{h_2}}}{(n+1)(n+2)}\right)\geq0.
\end{eqnarray}
The coefficients $a_n$ which are non-perturbative can be determined
from the initial condition $\phi_{\eta_c}(x_i,\mu_0)$ by using the
orthogonality relations for the Gegenbauer polynomials $C^{3/2}_n$.

\begin{figure}
\centering
\includegraphics[scale=1.20]{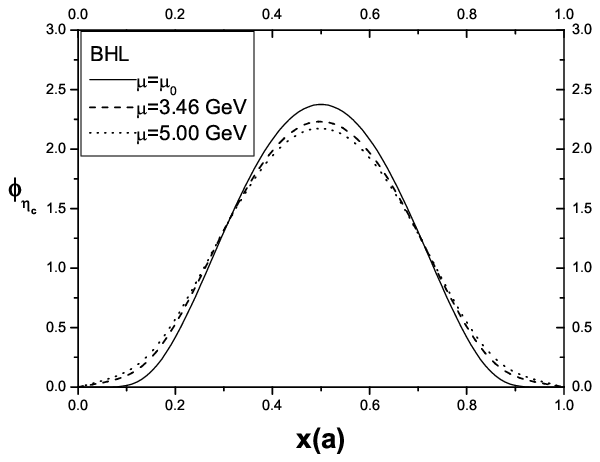}
\includegraphics[scale=1.20]{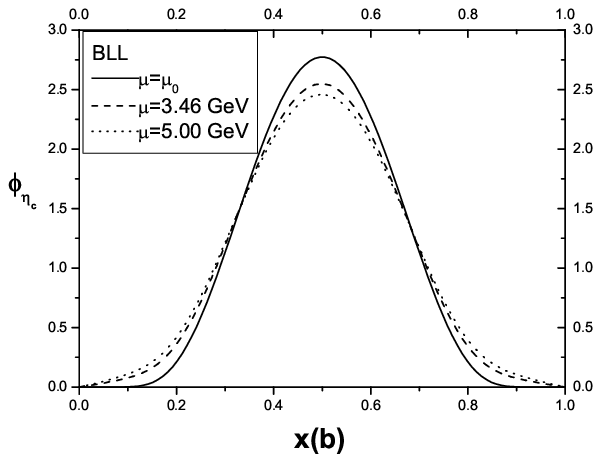}
\caption{$\eta_c$-DA derived with strict evolution
(\ref{eq:evolution}) at some typical energy scales, where the left
is for BHL model \protect \cite{BHL1981} and the right is for BLL
model \protect \cite{BLL2007}. The solid lines, dashed lines and the
dotted lines represent DA at $\mu=\mu_0$, $\mu=3.46~\mbox{GeV}$ and
$\mu=5.00~\mbox{GeV}$ respectively. }\label{fig2}
\end{figure}

Usually, one truncates the Gegenbauer expansion (\ref{eq:geg}) with
the first $3$ or $4$ terms ($n=0,2,4,6$ in our case) to obtain the
behavior of DA at the higher energy scales. In this paper we solve
the evolution equation Eq.(\ref{eq:evolution}) strictly to get the
DA's behavior at the large scale since DA's behavior is very
important for calculating the form factor of the process
$e^{+}+e^{-}\rightarrow J/\psi+\eta_c$. The Eq.(\ref{eq:evolution})
and Eq.(\ref{eq:geg}) are equivalent to each other if the Gegenbauer
expansion converges quickly. The evolution of DA with the strict
evolution (\ref{eq:evolution}) are shown in Fig.(\ref{fig2}), where
the solid lines represent the DAs at the initial energy scale
$\mu_0=1.80~\mbox GeV$, the dashed lines and dotted lines represent
the DAs at the energy scale $\mu=3.46~\mbox GeV$ and $\mu=5~\mbox
GeV$ that are taken by Ref.\cite{BC2005} and Ref.\cite{BLL2007}
respectively. It is shown that when the energy scale becomes larger,
the DA becomes lower in the middle while becomes higher near the end
point, till at last when the energy scale tends to infinity, the DA
tends to a asymptotic form $\phi_{as}(x)=6x(1-x)$.

\section{numerical results and discussion}

To calculate the form factor and the cross section of
$e^{+}+e^{-}\rightarrow J/\psi+\eta_c$, we take
$f_{J/\psi}=0.416~\mbox {GeV}$, $f_{\eta_c}=0.335~\mbox {GeV}$,
$m_{J/\psi}=3.096916~\mbox {GeV}$ and $m_{\eta_c}=2.9798~\mbox
{GeV}$ \cite{Ed2001,Am2008}. And to compare with the results in
literature \cite{BC2005}, we take the $c$-quark current mass to be
$m_c=1.2~\mbox {GeV}$. The Borel parameter $M^2$ ranges from 8 to 15
$\mbox{GeV}^2$, when at this range, both the form factor and the
cross section are stable. As for the effective scale $\mu$ of the
process $e^{+}+e^{-}\rightarrow J/\psi+\eta_c$, Ref.\cite{BC2005}
suggested $\mu \approx \sqrt{k^2} \approx 3.46~\mbox {GeV}$ from the
mean value of $<Z^k_m> \approx 0.80$ or from the coupling constant
$<\alpha_s(k^2)> \approx 0.263$. Another usually adopted scale is
$\mu \approx \sqrt{S}/2 \simeq 5~\mbox {GeV}$ \cite{BLL2007}. Here,
we will take $\mu=3.46~\mbox {GeV}$ and $\mu=5.00~\mbox {GeV}$ to do
our discussion.

\begin{figure}
\includegraphics[scale=1.20]{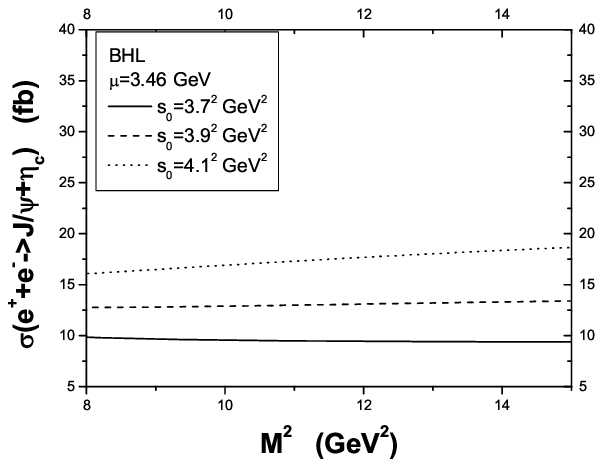}
\includegraphics[scale=1.20]{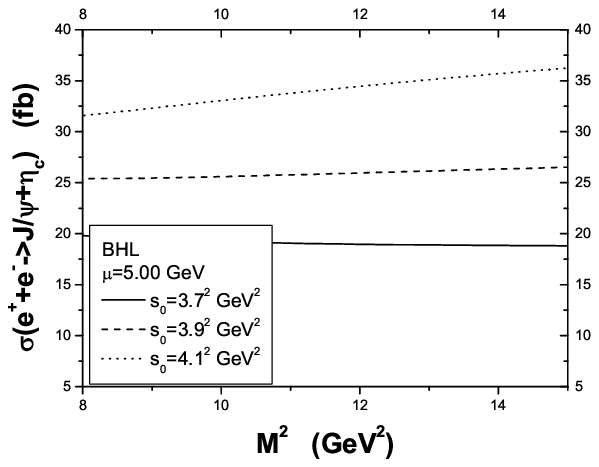}
\caption{The dependance of cross section on the threshold parameter
$s_0$ within the LCSR approach. The left and the right correspond to
the scale $\mu=3.46~\mbox{GeV}$ and $\mu=5.00~\mbox{GeV}$ with the
BHL model of $\eta_c$ DA.}\label{fig3}
\end{figure}

As for the threshold parameter $s_0$, Ref.\cite{EJ2001} took
$3.6^2~\mbox {GeV}^2<s_0<4.2^2~\mbox {GeV}^2$ \cite{EJ2001} with the
central value $s_0=3.8^2~\mbox {GeV}^2$ as their case. Similarly, we
also take $s_0$ within the same region while with a little different
central value. To see the dependence of the cross section on the
threshold parameter $s_0$, we give the cross section corresponding
to $s_0=3.7^2,\;\ 3.9^2,\;\ 4.1^2~\mbox GeV^2$ with the BHL model at
the scale $\mu=3.46~\mbox {GeV}$ and $\mu=5.00~\mbox {GeV}$ in
Fig.(\ref{fig3}). Since the cross section with the threshold
parameter $s_0=3.9^2~\mbox {GeV}^2$ is more stable than that with
$s_0=3.7^2~\mbox {GeV}^2, \;\ 4.1^2~\mbox {GeV}^2$, we take
$s_0=3.9^2~\mbox {GeV}^2$ as our central value of threshold
parameter.

\begin{figure}
\includegraphics[scale=1.20]{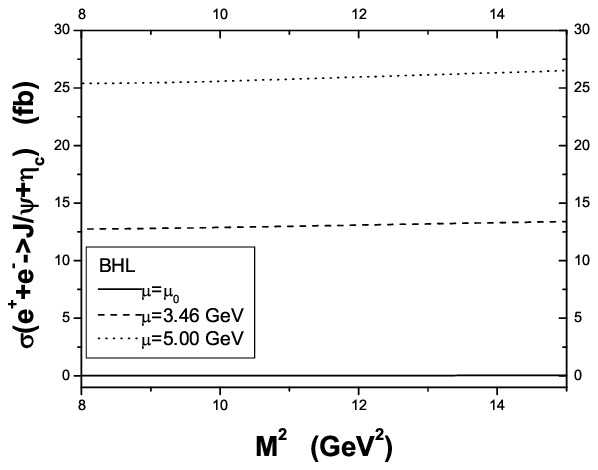}
\includegraphics[scale=1.20]{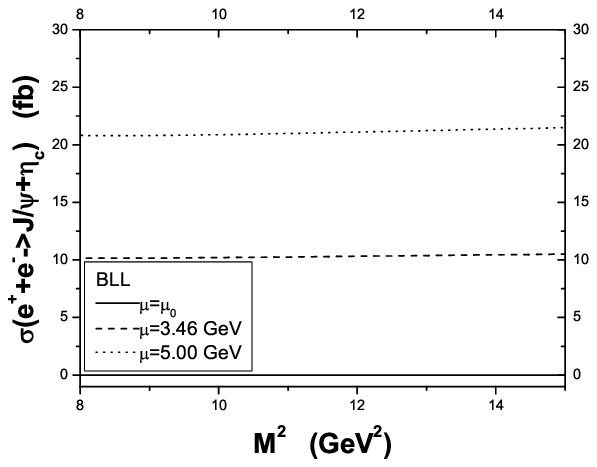}
\caption{The cross section of $e^{+}+e^{-}\rightarrow J/\psi+\eta_c$
at different effective scale within the LCSR approach. The solid
lines, dashed lines and dotted lines correspond to the $\eta_c$
meson distribution amplitude at the scale $\mu=\mu_0$,
$\mu=3.46~\mbox{GeV}$ and $\mu=5.00~\mbox{GeV}$, where the left is
for BHL model \protect \cite{BHL1981} and the right is for BLL model
\protect \cite{BLL2007}.}\label{fig4}
\end{figure}

\begin{table}
\begin{center}
\caption{Comparison of the cross section of $e^{+}+e^{-}\rightarrow
J/\psi+\eta_c$ from different $\eta_c$ DA models within the QCD
light-cone sum rules approach.}\label{tab2}
\begin{tabular}{|  c  |  c  |  c   |  c  |  c  |   c  |  c  |}
\hline $\phi$&BHL&BHL&BLL &BLL\\\hline
 $\mu~\mbox{(GeV)}$& 3.46 &5.00& 3.46 &5.00\\\hline
 $\sigma~\mbox{(\rm fb)}$&13.08$\pm$
0.32&25.96$\pm$0.55&10.34$\pm$0.17&21.16$\pm$0.34\\\hline
\end{tabular}
\end{center}
\end{table}

It is found that the cross section depends on the DA in the region
$x>\Delta$ through the form factor formula (\ref{eq:fvp}). We show
the cross section corresponding to three typical scales
$\mu=\mu_0=1.80~\mbox {GeV}$, $\mu=3.46~\mbox {GeV}$ and
$\mu=5.00~\mbox {GeV}$ in Fig.(\ref{fig4}). When the effective
energy scale increases, the corresponding cross section becomes
bigger and is compatible with the BaBar and Belle's measurements.
Thus, by setting the effective energy scale and dealing with the DA
evolution properly, the LCSR can provide a possible explanation for
the double-charmonium production process $e^{+}+e^{-}\rightarrow
J/\psi+\eta_c$ at the B factory. To explicate the cross sections of
different models numerically, we further show these cross sections
in Tab. \ref{tab2}, where the error is caused by the variation of
$M^2$. One may observe that the cross section by BLL model is
smaller than that by BHL model since the DA of the BLL model is
narrower than that of the BHL model as shown in Fig.(\ref{fig2}).

The above calculation is done at the LO approximation within the QCD
LCSR and the error is only caused by the variation of the Borel
parameter $M^2$. Of course, one should include the higher order
contributions, such as the NLO corrections to the light-cone sum
rules, the higher-twist DAs, the higher Fock states and etc.
Therefore we can not estimate all of the uncertainties of the
calculated cross section before doing a further study.

\section{summary}

The exclusive charmonium $J/\psi+\eta_c$ production in $e^{+}e^{-}$
collision is a very interesting problem. Since the discrepancy
between theoretical prediction of the LO NRQCD and experimental data
given by Belle and BaBar at the B factory has posed a significant
challenge for several years, many theoretical attempts have been
made to solve this challenging problem. It is worthwhile to study
this process by taking various applicable approaches to understand
the charmonium production dynamics. In this paper we study this
process by using the QCD LCSR approach.

Our results based on the LCSR approach shows that the cross section
of the process $e^{+}+e^{-}\rightarrow J/\psi+\eta_c$ substantially
depends on the behavior of the $\eta_c$ DA at the energy scale
$\mu$. Noticing that the energy scale $\mu$ at the B factory is
greater than the initial scale $\mu_0$ of the $\eta_c$ DA, the
renormalization group evolution of the DA has to be taken into
account. The perturbative radiative correction leads to a big change
of the DA especially to the tail of the DA at the large scale $\mu$.

At the present, one has poor knowledge of the DA and tries to build
various models that have quite different behavior, especially at the
end-point region. We stress that the evolution of the $\eta_c$ DA
can give more reasonable prediction to the process
$e^{+}+e^{-}\rightarrow J/\psi+\eta_c$ within the LCSR approach.
Similar to other approaches, in order to calculate the cross section
of the double-charmonium production one needs to have more knowledge
of the charmonium DA.

The numerical results show that the cross section of the process
$e^{+}+e^{-}\rightarrow J/\psi+\eta_c$ is predicted in the range
$13\sim 26 fb$. The calculated values for the different models can
be compatible with the Belle and BaBar measurements by properly
choosing effective energy scale for this process and dealing with
the DA evolution effect.

\hspace{1cm}

{\bf Acknowledgments}: This work was supported in part by Natural
Science Foundation of China under Grant No.10675132, No.10735080 and
No.10805082, and by Natural Science Foundation Project of CQ CSTC
under Grant No.2008BB0298.


\begin{thebibliography}{99}

\bibitem{Belle2002}K, Abe, {\it et al.}, Phys. Rev. Lett. {\bf 89}, 142001 (2002).

\bibitem{Belle2004}K, Abe, {\it et al.}, Phys. Rev. D {\bf 70}, 071102 (2004).

\bibitem{BaBar2005}B. Aubert, Phys. Rev. D {\bf 72}, 031101 (2005).

\bibitem{BBL1997} G.T. Bodwin, E. Braaten and G.P. Lepage,
Phys. Rev. D {\bf 51}, 1125 (1995); Erratum Phys. Rev. D {\bf 55},
5853 (1997).

\bibitem{BLB2003}G.T. Bodwin, J. Lee and E. Braaten, Phys. Rev. Lett. {\bf 90}, 162001(
2003).
\bibitem{BL2003}E. Braaten and J. Lee, Phys. Rev. D {\bf 67}, 054023
(2003).

\bibitem{Chao2003}K.Y. Liu, Z.G. He, K.T. Chao, Phys. Lett. B {\bf 557}, 45
(2003).

\bibitem{Chao2006}Y.J. Zhang, Y.J. Gao and K.T. Chao, Phys. Rev. Lett. {\bf 96}, 092001
(2006).

\bibitem{Wang2008} B. Gong and J.X. Wang, Phys.Rev. D {\bf 77}, 054028(2008);
Phys. Rev. Lett. {\bf 100}, 181803(2008).

\bibitem{BKKY2007}G.T. Bodwin, D. Kang, T. Kim, J. Lee and C. Yu, AIP Conf. Proc. {\bf
892}, 315(2007).

\bibitem{Chao2007}Z.G. He, Y. Fan, K.T. Chao, Phys. Rev. D {\bf 75}, 074011
(2007).

\bibitem{BLC2008}G.T. Bodwin, J. Lee, C. Yu, Phys. Rev. D {\bf 77}, 094018
(2008).

\bibitem{BC2005}A.E. Bondar, V.L. Chernyak, Phys. Lett. B {\bf 612}, 215
(2005).

\bibitem{K1999}A. Khodjamirian, Eur. Phys. J. C {\bf 6}, 477 (1999).

\bibitem{Braguta2008}V.V. Braguta, PoS Confinement {\bf 8}, 097 (2008),
hep-ph/08112640.

\bibitem{CJ2007}H.M. Choi, C.R. Ji, Phys. Rev. D {\bf 76}, 094010 (2007).

\bibitem{BBK1989}I.I. Balitsky, V.M. Braun, A.V. Kolesnichenko, Nucl. Phys. B {\bf 312}, 509
(1989).

\bibitem{BF1989}V.M. Braun, I.E. Filyanov, Z. Phys. C {\bf 44}, 157 (1989).

\bibitem{CZ1990}V.L. Chernyak, I.R. Zhitnitsky, Nucl. Phys. B {\bf 345}, 137
(1990).

\bibitem{CK2001}P. Colangelo, A. Khodjamirian, ``QCD Sum Ruless, a Modern Perspective",
hep-ph/0010175; Boris Ioffe Festschrift`` At the Frontier of
Particle Physics / Handbook of QCD'', edited by M. Shifman (World
Scientific, Singapore, 2001).

\bibitem{EJ2001}M.E idemuller, M. Jamin, Phys.Lett.B {\bf 498}, 203(2001);
Nucl. Phys. Proc. Suppl. {\bf 96}, 404 (2001).

\bibitem{SVZ1979}M.A. Shifman, A.I. Vainshtein, V.I. Zakharov, Nucl. Phys. B {\bf 147}, 385, 448
(1979).

\bibitem{LB1980}G.P. Lepage, S.J. Brodsky, Phys. Rev. D {\bf 22}, 2157
(1980).

\bibitem{BKL2006}G.T. Bodwin, D. Kang, J. Lee, Phys. Rev. D {\bf 74}, 114028(2006).

\bibitem{BLL2007}V.V. Braguta, A.K. Likhoded, A.V. Luchinsky, Phys. Lett. B {\bf 646}, 80
(2007).

\bibitem{MS2004}J.P. Ma, Z.G. Si, Phys. Rev. D {\bf 70}, 074007 (2004).

\bibitem{BHL1981} S. J. Brodsky, T. Huang and G. P. Lepage, in {\it
Particles and Fields-2}, Proceedings of the Banff Summer Institute,
Banff, Alberta, 1981, edited by A. Z. Capri and A. N. Kamal (Plenum,
New York, 1983), p143; G. P. Lepage, S. J. Brodsky, T. Huang, and P.
B. Mackenize, {\it ibid.}, p83; T. Huang, {\it in Proceedings of
XXth International Conference on High Energy Physics}, Madison,
Wisconsin, 1980, edited by L. Durand and L. G. Pondrom, AIP Conf.
Proc. No. 69 (AIP, New York, 1981), p1000.

\bibitem{HZ2007}T. Huang, F. Zuo, Eur. Phys. J. C {\bf 51}, 833 (2007).

\bibitem{M1974}H.J. Melosh, Phys.Rev. D{\bf 9}, 1095(1974).

\bibitem{spin1}T. Huang, B.Q. Ma and Q.X. Shen, Phys.Rev.D {\bf 49},
1490(1994).

\bibitem{spin2} F.G. Cao, T. Huang, Phys. Rev. D {\bf 59}, 093004(1999).

\bibitem{spin3} T. Huang, X.G. Wu and X.H. Wu, Phys.Rev. D{\bf 70},
093013(2004); T. Huang and X.G. Wu, Int. J. Mod. Phys. A{\bf 22},
3065(2007)

\bibitem{Yao2006}Y.M. Yao {\it et al.}, Partilce Data Group, J. Phys. G {\bf 33}, 1(2006).

\bibitem{Ed2001}K.W. Edwards {\it et al.}, Phys. Rev. Lett. {\bf 86}, 30(2001).

\bibitem{Am2008}C. Amsler et al., Particle Data Group, Phys. Lett.
B{\bf667}, 1 (2008).

\end{thebibliography}
\end{document}